# Temperature dependence of the anharmonic decay of optical phonons in carbon nanotubes and graphite

Ioannis Chatzakis,[1,2,*] Hugen Yan,[1] Daohua Song,[1] Stéphane Berciaud,[1,3]

and Tony F. Heinz[1]

[1]*Departments of Physics and Electrical Engineering, Columbia University, 538 West 120th St., New York, NY 10027*

[2]*Department of Physics, JRM Laboratory, Kansas State University Manhattan, KS 66506*

[3]*IPCMS (UMR 7504), Université de Strasbourg and CNRS, F-67034 Strasbourg, France*

## Abstract

We report on the temperature dependence of the anharmonic decay rate of zone-center (G-mode) optical phonons in both single-walled carbon nanotubes and graphite. The measurements are performed using a pump-probe Raman scattering scheme with femtosecond laser pulses. For nanotubes, measured over a temperature range of 6 K – 700 K, we observe little temperature dependence of the decay rate below room temperature. Above 300 K, the decay rate increases from 0.8 ps$^{-1}$ to 1.7 ps$^{-1}$. The decay rates observed for graphite range from 0.5 to 0.8 ps$^{-1}$ for temperatures from 300 – 700 K. We compare the behavior observed in carbon nanotubes and graphite and discuss the implications of our results for the mechanism of the anharmonic decay of optical phonons in both systems.

# I INTRODUCTION

Single-walled carbon nanotubes (SWNTs) and graphene, with their exceptional current-carrying capabilities, are materials of fundamental interest, as well as potential candidates for future electronic devices [1]. In the regime of high-field transport, carriers with excess kinetic energy can emit high-energy optical phonons. These optical phonons decay into lower energy phonons through anharmonic coupling [2-5]. If the anharmonic decay rate of the optical phonons is not sufficiently rapid a bottleneck can be created. Optical phonons may then accumulate and build up a non-equilibrium population. Such "hot" phonons have been invoked to explain the observation of negative differential conductance in suspended carbon nanotube field-effect transistors under conditions of high electrical bias [2, 4, 6]. The anharmonic decay rate of the optical phonons is a critical parameter in modeling of high-field transport. To understand and, possibly, engineer energy flow in these devices, we wish to obtain a quantitative description of the decay channels of these optical phonons.

Experimentally such phonon-phonon interactions can be examined spectroscopically by different approaches. Under appropriate conditions the line widths of the vibrational transition, observed either by infrared or Raman spectroscopy, provide information on the anharmonic decay rate of the relevant phonons. These frequency-domain methods require, however, the lack of inhomogeneous broadening, something that is often difficult to achieve in nanoscale materials. Time-domain pump-probe techniques [7] have the advantage of providing direct access to phonon lifetimes even in the presence of inhomogeneous line broadening. In this approach, non-equilibrium phonon populations are excited directly or through electronic decay and the evolution of the population is followed either by pulsed infrared radiation or by optical radiation through the Raman process. The combination of vibration excitation via an ultrafast optical pulse and time-resolved Raman scattering [7] has recently been successfully applied to examine the dynamics of optical phonons in both carbon nanotubes and graphite [8-10]. For systems such as these that may exhibit strong electron-phonon coupling, the pump-probe time-domain method offers a further significant advantage in determining anharmonic decay rates. The approach circumvents complications associated with the electronic contribution to the phonon line width [11] and provides, as we discuss below, a direct measurement of the anharmonic decay rate, *i.e.*, the decay rate of the optical phonons into other lower energy phonons. Our work complements the study of Ishioka et al. [12], who examined the G-mode coherent phonons dynamics of graphite by means of transient reflectivity. Coherent phonon spectroscopy has also been applied to carbon nanotubes [13, 14], and coupling between G mode and radial breathing mode phonons was revealed. Recently, Ikeda et al. [15] studied the phonon dynamics in SWNTs using broadband coherent anti-Stokes Raman scattering microscopy in time-frequency domain. They suggested that the pure dephasing contribution to the line width is negligible at room temperature. Importantly, a measurement of the temperature dependence of the anharmonic decay rate can yield quantitative results about the available decay channels. To date, such detailed information about phonon-phonon couplings, although investigated theoretically [11, 16, 17], has not been available from experiment.

In this paper, we examine the temperature dependence of the anharmonic decay rate in both highly-oriented pyrolitic graphite (HOPG) and SWNTs. The zone-center (G-mode, ~1585 cm$^{-1}$) optical phonon lifetime was measured from cryogenic temperatures to 700 K by means of time-resolved pump-probe Raman spectroscopy. For both materials, we observed that the decay rate increases with temperature, with SWNTs exhibiting a more pronounced temperature dependence than HOPG. The temperature dependence of the anharmonic decay rates provides insight into the decay channels for such decay process. In particular, since the decay rate is enhanced by the population of the daughter modes, the temperature dependence of the decay yields, through the thermal activation characteristics, information about the energy of the daughter modes. In this work, we analyze the decay based on three-phonon processes to draw conclusions about the energies of the phonons produced by anharmonic decay. We also compare our results with existing *ab-initio* simulations [11]. Our results for the temperature dependence of the decay rate in graphite are in good agreement with theoretical predictions for graphite, which are in turn based on modeling of a single sheet of graphene [11]. On the other hand, these theoretical predictions [11] significantly underestimate the decay rate observed for SWNTs and also fail to replicate the observed temperature dependence of the rate. These findings suggest the existence of additional channels for anharmonic decay that are inherent to carbon nanotubes and are not present in graphene.

**II EXPERIMENTAL**

In our experiments, an ultrashort pump laser pulse was used to generate a non-equilibrium population of optical phonons. The phonons are generated by the rapid relaxation of the photoexcited carriers[18-20]. The dynamics of the resulting hot phonons was then measured by recording the G-mode Raman signal generated by means of a time-delayed probe pulse. Since the intensity of the anti-Stokes G-mode feature is proportional to the population of optical phonons involved, one can directly monitor the phonon decay by varying the delay between the ultrafast pump and probe pulses [8,9].

The pump beam consisted of laser pulses with an optimal duration of ~100 fs at a photon energy of 1.55 eV that were supplied by a 1-kHz repetition rate, regeneratively amplified, mode-locked Ti:sapphire laser. The probe pulses of 3.1 eV photon energy were obtained by frequency doubling of the pump beam. The probe pulses had a slightly longer duration (~150 fs under optimal conditions) due to the dispersion of the band-pass filters used to block the residual pump photons. In some of the measurements, laser instability causes broadening of the pulses by up to a factor of 2. The pump and probe pulses were introduced in a nearly collinear geometry, the delay between them varied using a motorized translation stage. The two beams had identical, linear polarizations. A charge-coupled device (CCD) cooled by liquid nitrogen served to detect the Raman-scattered photons from the probe beam after dispersal of the radiation in a grating spectrometer. In the pump-probe measurements, typical fluences were 2 mJ/cm$^2$ for the pump and 0.7 mJ/cm$^2$ for the probe pulses. The phonon decay rate was examined at somewhat higher and lower pump and probe fluences. Within the accuracy of the measurement, we found no significant fluence dependence. For low-temperature measurements, the samples were mounted on the cold finger of a cryostat equipped with optical windows. For high temperatures, a homemade electrical heater was employed. In all cases, the sample temperature was determined with a type-K thermocouple. All measurements were carried out under vacuum (10$^{-6}$ Torr).

The SWNT samples were prepared as a film by dissolving commercially obtained nanotubes in dichloroethane and allowing the solution to evaporate on a fused-quartz substrate. The nanotubes were synthesized by the high-pressure carbon monoxide (HiPCO) process (purchased from CheapNanotubes), which produced high purity semiconducting and metallic SWNTs with 1-2 nm diameter and 5-30 µm length. The sample consists largely of nanotube bundles. The 1.55 eV pump photons resonantly excite the $S_{22}$ transition of semiconducting species with diameters in the range 1.1-1.2 nm, as well as the $M_{11}$ transition of metallic species with diameters of ~1.7 nm [21-23]. For the graphite sample, we used HOPG with a mosaic spread of 0.8º (purchased from Advanced Ceramics). We note with respect to the Raman measurements that in nanotubes (but not in graphite), there are actually two distinct G-mode phonon frequencies, corresponding to phonons polarized in the transverse (TO phonons) and longitudinal (LO phonons) directions with respect to the nanotube axis. Because of the significant bandwidth (~ 196 cm$^{-1}$) of our ultrafast probe pulses, the observed Raman spectra in our time-resolved measurements do not permit the identification of separate LO and TO features (Fig. 1). The measured spectrally integrated response is presumably dominated by the LO phonons, which generally contribute more strongly to the Raman scattering process [24-26].

**III RESULTS**

The basis for our time-resolved Raman measurement can be understood readily from the anti-Stokes Raman spectra obtained from the probe beam alone, as well as from the probe beam with simultaneous excitation by the pump beam. Typical experimental data are shown in Fig. 2 for SWNTs at different temperatures. The anti-Stokes signal, which scales linearly with the phonon population *n*, is very weak for the probe beam alone. The signal increases by about 1 order of magnitude through of an appreciable G-mode phonon population. The width of the Raman feature does not reflect the inherent linewidth of the G-mode vibration, but rather the large bandwidth of the ultrashort probe laser pulse. A slight blue shift in G-mode frequency can also be observed in the pumped spectrum. This effect, which arises from coupling of the vibration to the nonequilbrium electronic excitation [9], is not of importance to our current investigation.

The temporal evolution of the (spectrally integrated) G-mode anti-Stokes Raman intensity is shown in Fig. 2 for both the HOPG and SWNT samples over a range of samples temperatures $T$. We interpret this signal as recording the population dynamics. As reported earlier [9], we see a steep rise in the phonon population during excitation of the samples by the pump pulse. This response is attributed to the strong coupling between the photoexcited carriers and the optical phonons, leading to the creation of a nonequilibrium phonon population. The initial rise in phonon population occurs on the time scale of a few hundred femtoseconds and is dominated by our instrumental response time. The subsequent decay of the phonon population on the time scale of a picosecond is the subject of this study.

The temporal evolution of the Raman data for both SWNTs and HOPG can be fit (Fig. 2, solid lines) to an abrupt rise followed by a single-exponential decay, once the finite instrumental response time is taken into account by convolution with a Gaussian envelope. In our data, we subtract the weak anti-Stokes signal present without the pump, *i.e.*, for negative time delays. We denote the corresponding decay rate for the phonon population as $\Gamma_{ph-ph}$. The variation of $\Gamma_{ph-ph}$ with the sample temperature $T$ extracted from these data and additional measurements (not shown) is summarized in Fig. 3. We also express the decay rate $\Gamma_{ph-ph}$ in terms of the corresponding lifetime broadening of the vibrational line, measured as the full-width at half maximum (FWHM). In wave numbers, this quantity is given by $\Gamma_{ph-ph}/2\pi c$.

## IV DISCUSSION

### (a) Comparison of room-temperature data

Let us first consider the behavior observed in the measurements performed at room temperature. Our results for SWNTs are consistent with previous studies based on time-resolved Raman scattering [8, 9, 27]. We note that in the case of SWNTs, the measured phonon decay rates reflect the response from an ensemble of metallic and semiconducting nanotubes. The good agreement with earlier studies [8, 9], despite the use of different substrates and average nanotube diameters, confirms that the decay rates of the optical phonons do not depend critically on the local environment or on the SWNTs structural parameters.

The anharmonic decay rate observed for graphite in the present experiment (0.52 ± 0.02 ps$^{-1}$ or 2.76 ± 0.04 cm$^{-1}$) at 340 K is compatible with the value (0.45± 0.02 ps$^{-1}$ or 2.41 ± 0.10 cm$^{-1}$) reported earlier for a pump-probe measurement of graphite at room-temperature [9]. A further immediate finding is that the phonon decay rate in the nanotube samples (0.90 ±0.09 ps$^{-1}$ or a lifetime broadened FWHM of 4.8 ±0.5 cm$^{-1}$) is significantly greater than that observed for bulk graphite (0.45 ± 0.02 ps$^{-1}$ or 2.41 ± 0.1 cm$^{-1}$) [9].

It is also interesting to compare our decay rates with those implied by the line widths of conventional Raman scattering. Such frequency-domain measurements may exhibit broader line widths because of inhomogeneous broadening effects when probing ensemble samples or even the inhomogeneous environment experienced by an individual nanotube. The Raman line widths for ensemble samples of SWNTs, typically lie in the range of 15 cm$^{-1}$, far exceeding the lifetime broadening of 4.8 cm$^{-1}$ implied by our time domain measurements. A better comparison is provided by measurements of individual nanotubes. For semiconducting nanotubes, significantly narrow linewidths can be obtained. Due to the existence of sizable electronic energy gap, the optical phonons can not efficiently couple to electron-hole pairs. Thus, the electron-phonon coupling almost has no contribution to the linewidth [28]. The Raman width in individual semiconducting SWNTs at 300 K is typically 5-10 cm$^{-1}$ [29], which is very close to ~5 cm$^{-1}$ deduced from our time resolved data. The slight difference between the two might be due to inhomogeneous broadening. However, for individual metallic SWNTs, the observed G-mode width is generally 10's of cm$^{-1}$ [24-26, 28]. Similarly, for conventional frequency-domain Raman measurements in HOPG, the width of the G-mode was found to be close to 15 cm$^{-1}$ [28], compared to < 3 cm$^{-1}$ of lifetime broadening in the time-domain measurements reported here. These discrepancies with the time-domain measurements reflect the decay of phonons through electron-hole pair creation. As will be discussed below, this channel does not significantly influence the decay rate observed in our time-domain pump-probe study. However, it can be the dominant effect in determining the Raman line width $\Gamma = 1/(\pi c T_2)$, where $2/T_2 = 1/\tau + 1/T_1$ and $\tau$ the pure dephasing time, in conventional spectroscopic measurements. The incoherent time-resolved anti-Stokes Raman scattering that is applied in the present work yields access to $T_1$.

**(b) Temperature dependence of the anharmonic decay rate**

The temperature dependence of the phonon decay rates in both the SWNTs and HOPG shows a similar overall trend. There is no significant temperature dependence for SWNTs below room temperature, while the temperature dependence of HOPG near room temperature is relatively weak. The detailed decay rates and their variation with temperature are, however, different for the two systems. In addition to the greater absolute phonon decay rate in SWNTs than in graphite at room temperature, the increase of the decay rate with temperature is considerably more pronounced in SWNTs.

Before presenting a more extensive analysis of the implication of the phonon decay rates and their temperature dependence, we comment briefly on the interpretation of the measured decay rate, *i.e.*, the temporal evolution of the signal. What processes contribute to this measured decay? The channel common to all materials is one involving coupling to other phonons, *i.e.*, the anharmonic decay process. In addition, optical phonons in metallic nanotubes and graphite are strongly coupled to low-energy electronic excitations. We now argue that this coupling does not significantly influence the measured decay rate in our pump-probe measurement, which can, consequently, be interpreted as directly providing the anharmonic phonon decay rate. As has been discussed in previous papers [8, 9] for the case of metallic SWNTs and graphite, after excitation by the ultrafast pump laser pulse, the electronic system quickly comes into and remains in thermal equilibrium with the optical phonons. This process leads to the rapid increase observed in the optical phonon population following photoexcitation. The measured decay can then be regarded as that of a subsystem consisting of the phonons coupled to the electronic excitations. However, since the electronic excitations also have much lower heat capacity than the excited optical phonons, the decay of the phonons is not significantly influenced by coupling to the electronic excitations [9]. We can thus consider the measured decay dynamics for the phonon population to reflect anharmonic decay rates. We note that for the case of the nanotubes, with their strong contact to the external environment, decay into vibrations in the external media may in principle also be relevant. This channel is, however, not considered to play a significant role in the decay dynamics, which appear to be dominated by intrinsic decay channels [8].

**(c) Microscopic models for the decay process**

We now consider the temperature dependence of the measured anharmonic decay rate of the optical phonons on the microscopic level. The lowest-order anharmonic coupling corresponds to the third-order term in the expansion of the potential energy of the crystal. This coupling gives rise to a decay of the optical phonon into two daughter phonons. If we denote the corresponding frequencies and wave vectors, respectively, by $\omega_1$, $\omega_2$ and $\mathbf{q}_1$, $\mathbf{q}_2$, then conservation of energy and momentum requires $\hbar\omega_1 + \hbar\omega_2 = \hbar\omega_G$ and $\mathbf{q}_1 + \mathbf{q}_2 = 0$, where $\omega_G = 1585 \text{ cm}^{-1}$ is the G-mode frequency. The temperature dependence of this process arises from the fact that the anharmonic decay process is stimulated by the presence of population in the daughter phonon modes according to [16]

$$\Gamma_{ph-ph}(T) = \Gamma_0 \left[ 1 + n(\hbar\omega_1, T) + n(\hbar\omega_2, T) \right]. \tag{1}$$

Here $\Gamma_0$ denotes the decay rate at zero temperature, and $n(\hbar\omega_1, T)$ and $n(\hbar\omega_2, T)$ are the populations of the two daughter modes for the decay process. This expression is written for a single dominant decay mode, but can obviously be generalized to include additional channels. Assuming that the pump-induced perturbation of the sample is weak, in accordance with the lack of a significant dependence of the decay rates on the fluences of the pump pulses, we can use Bose-Einstein statistics to determine the populations $n(\hbar\omega_1,T)$ and $n(\hbar\omega_2,T)$ of the secondary phonons modes. We see that analysis of the temperature dependence of the phonon lifetime thus provides information about the dominant decay channels. We note that in the above expression, we exclude contributions from phonon up-conversion processes, since the G-mode phonons correspond to the highest-energy branch [11].

We first compare our experimental findings with the decay channels predicted by Bonini *et al.* using *ab-initio* simulations of the behavior in graphene [11]. To our knowledge, there are no available theoretical predictions for the anharmonic decay channels calculated specifically for SWNTs or bulk graphite, although it is argued by Bonini *et al.* [11] that the behavior in nanotubes can be approximated by that of graphene. Considering the dispersion of the six phonon branches in graphene, two classes of 3-phonon processes have been predicted to dominate the anharmonic decay. The first involves a production of a transverse acoustic (TA) phonon with a longitudinal acoustic (LA), with the former having a frequency of $\omega_1 = \omega_G - \omega_2$ in the range of 550 to 650 cm$^{-1}$. This channel is predicted to contribute 56% of the anharmonic decay rate. The second class is a symmetrical decay process involving two phonons of equal, but opposite momenta within the same branch (LA-LA, TA-TA, and ZO-ZO, where ZO refers to out-of-plane optical phonons). This channel, corresponding to the commonly used Klemens model [16] of symmetrical decay, is predicted to be responsible for the remaining 44% of the anharmonic decay rate. As may be expected from this description of the decay process, we find that the temperature dependence obtained in the *ab-initio* simulations of the anharmonic decay rate [11] can be fit well using the form of Eqn. (1) by including two indicated decay channels with an appropriate weighting, namely,

$$\Gamma_{ph-ph}(T) = \Gamma_0 \left[ 0.56 \times \left(1 + n(\hbar\omega_1, T) + n(\hbar\omega_G - \hbar\omega_1, T)\right) + 0.44 \times \left(1 + 2n(\hbar\omega_G/2, T)\right) \right],$$

(2)

with $\omega_1 = \omega_{1,th} = 630$ cm$^{-1}$ and $\Gamma_0 = \Gamma_{0,th} = 1.8$ cm$^{-1}$.

In Fig. 3, we compare this theoretical prediction for the temperature-dependent anharmonic decay rate with our experimental data for the two different material systems. For both cases, we find that the calculated low-temperature anharmonic decay rate of $\Gamma_{0,th} = 1.8$ cm$^{-1}$ underestimates the low-temperature limits observed in our experiment of $\Gamma_{ph-ph}(T \to 0) = \Gamma_0 = 2.5$ cm$^{-1}$ and 4.2 cm$^{-1}$ for HOPG and SWNTs, respectively. The comparison of absolute decay rate is not very informative. So we compare the predicted temperature dependence, which reflects the energies of the daughter phonons in the predicted decay channels and their relative strength, with the experiment by adjusting the low-temperature limit to match the experimental result. We obtain good agreement between experiment and the scaled theory for HOPG. This suggests that the weakness of interlayer interactions in graphite renders graphene a suitable model for calculation of the rate of the anharmonic decay of optical phonons. For phonons in SWNTs, however, the experimental temperature dependence is stronger than the predicted variation. This difference, as well as the substantially higher overall anharmonic decay rate for SWNTs suggests that additional decay channels are present in carbon nanotubes that are not captured in the modeling of graphene.

In order to obtain further insight into the dominant processes involved in the anharmonic decay, we have performed a free fit of our temperature dependent phonon decay rate to the simplified form of Eqn. (1), corresponding to an anharmonic decay with a single dominant channel. The parameters of our best fits are presented in Table 1. For the case of HOPG, we find that the secondary phonon energies inferred from our fit ($\omega_1 = 535$ cm$^{-1}$, $\omega_2 = 1050$ cm$^{-1}$) are quite close to the predictions of Bonini *et al.* for the TA-LA decay, as expected based on the agreement of their predicted temperature dependence with the experiment. Given the experimental signal-to-noise ratio, the single channel TA-LA decay and the combination of the two decay channels predicted by Bonini *et al.* describe the results equally well (Fig. 3). For the case of SWNTs, however, the best fit of the experimental results to Eqn. (1) yields phonon frequencies of $\omega_1 = 330$ cm$^{-1}$ and $\omega_2 = 1255$ cm$^{-1}$. The large energy difference between this secondary phonon pair cannot be described in terms of a process involving the phonon branches in graphene. We presume that the observed behavior reflects the contributions of additional modes that are present only in nanotubes, such as the radial breath mode (RBM) and the low-energy pinch and bending modes [30]. The influence of these decay channels presumably also leads to the significantly higher overall anharmonic decay rate in nanotubes compared with graphite. We note that the true phonon band structure of nanotubes, because of the circumferential quantization condition, actually departs from that of graphene and offers, at least in principle, many more decay processes that conserve energy and momentum. We hope that these results will trigger further theoretical studies of the channels for anharmonic decay in these prototypical systems of graphitic carbon.

## V CONCLUSIONS

In summary, we have directly measured the temperature dependence of the anharmonic decay of the zone-center (G-mode) optical phonons in graphite and single-walled carbon nanotubes by means of time-resolved anti-Stokes Raman spectroscopy. We find that the optical phonon decay rates increase from 0.52 ps$^{-1}$ to 0.79 ps$^{-1}$ over the temperature range of 340 K - 685 K for graphite, and from 0.81 ps$^{-1}$ to 1.66 ps$^{-1}$ over the temperature range of 6 K to 700 K for SWNTs. The temperature dependence of the decay rate in graphite agrees well with *ab-initio* simulations for graphene that involve a major TA-LA acoustic phonon decay channel [11]. The decay rate in SWNTs, however, increases considerably more rapidly with temperature than theory predicts for graphene. This suggests the presence of additional decay channels involving low frequency modes, such as the RBM, pinch and bending modes. With respect to the role of non-equilibrium populations of strongly-coupled optical phonons in nanotubes under conditions of high electrical bias, the present results suggest that the overpopulation of these modes may be smaller than previously anticipated based on the theoretically predicted (lower) anharmonic decay rate. The higher anharmonic decay rates of these phonons measured in this paper, particularly at elevated temperatures, will act to suppress the departure from equilibrium behavior.

This work was supported by the Office of Basic Energy Sciences of the U.S. Department of Energy and by Multidisciplinary University Research Initiative (MURI) of the Air Force Office of Scientific Research.

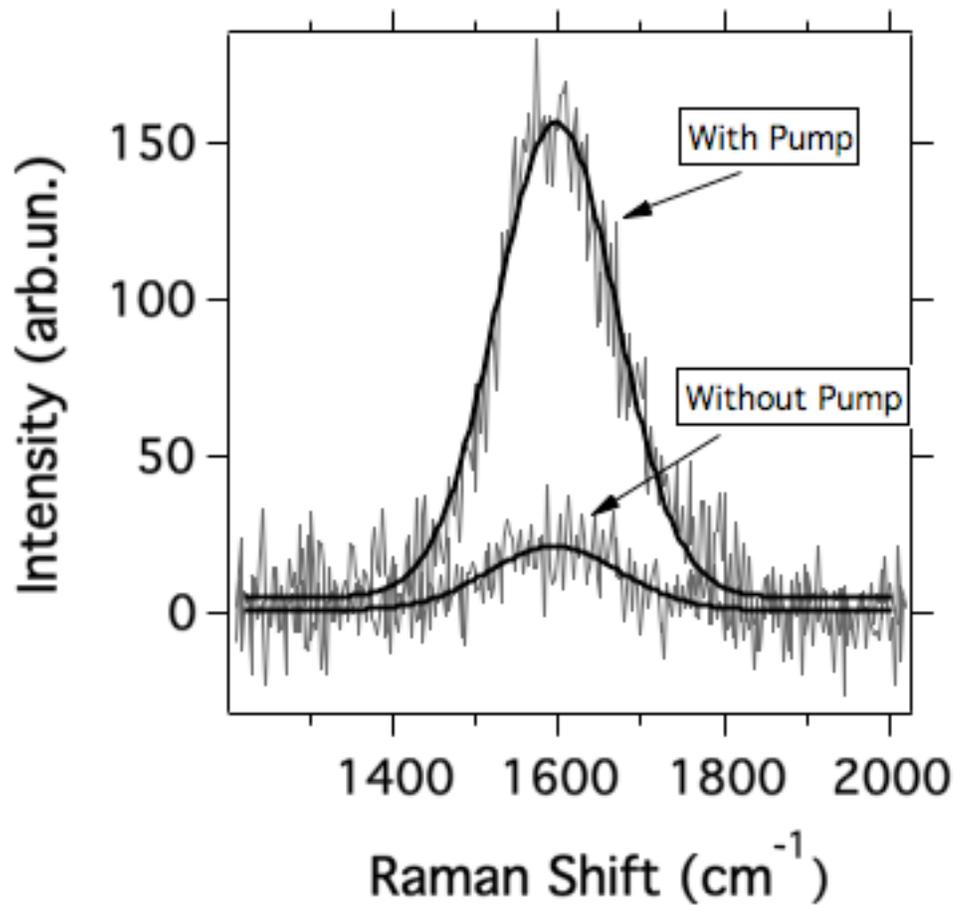

**Fig. 1:** Anti-Stokes Raman spectrum of G-mode in SWNTs at room temperature. The spectra are recorded with the spectrally broad, ultrafast probe pulse and are fit (solid lines) to a Gaussian peak. The results are shown both in the presence and absence of pump radiation.

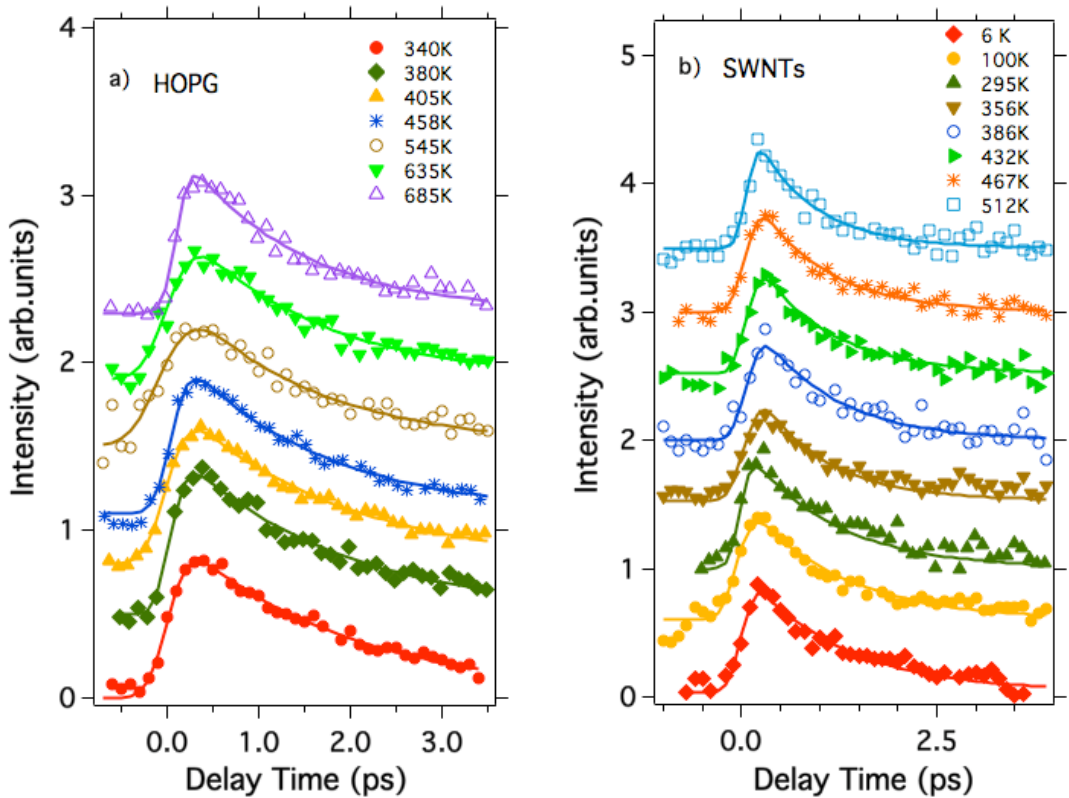

**Fig. 2:** (Color online) Time-resolved anti-Stokes Raman intensity for G-mode phonons in (a) HOPG and (b) SWNTs. The signal (symbols) is plotted as a function of the delay time between the pump (1.55 eV) and probe (3.1 eV) laser pulses for different temperatures. The solid lines are fits based on an abrupt rise in the phonon population followed by a single exponential decay, after convolving with the instrumental response time.

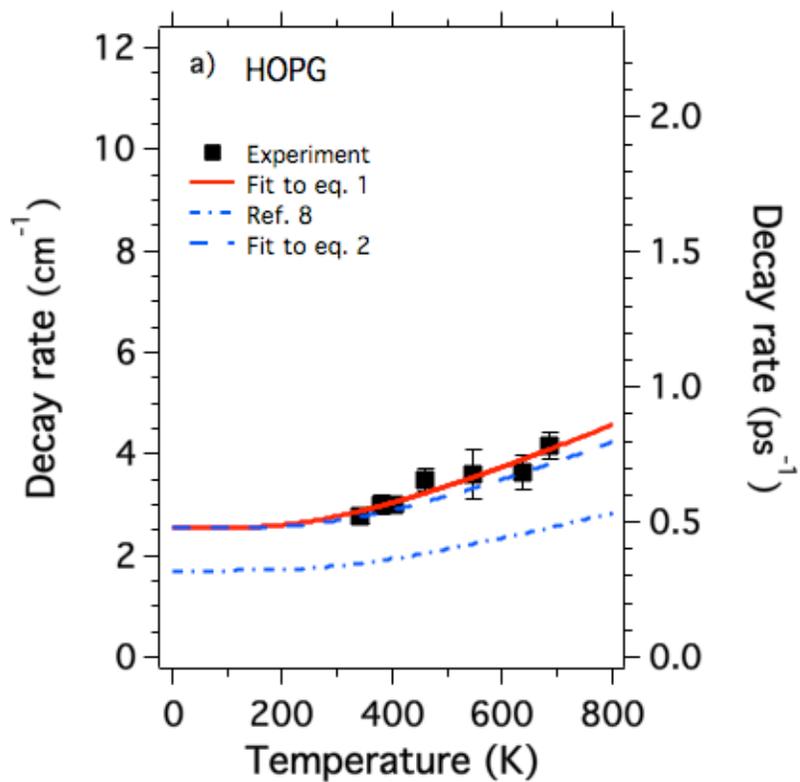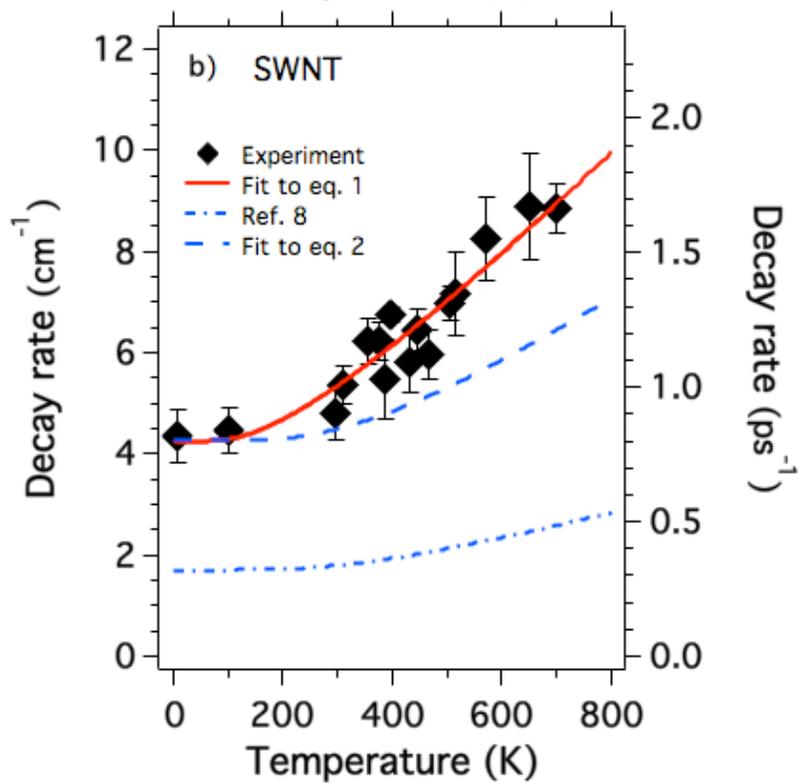

**Fig. 3:** (Color online) Decay rate (expressed in ps$^{-1}$ and cm$^{-1}$) of the G-mode optical phonons in (a) HOPG and (b) SWNTs as a function of temperature. The experimental data (symbols) are fit to Eqn. (1) (thick solid lines). The dashed lines represent the theoretical prediction of Bonini *et al.* [11], as summarized by Eqn. (2) in the text, for the predicted optical phonon decay rates. The predictions are shown with (dashed line) and without (dot-dashed line) a multiplicative correction factor of 1.5 and 2.5, respectively, for HOPG and SWNTs.

**Table 1**. Summary of the fit parameters using Eqn. (1) to describe the experimental temperature dependence of the anharmonic decay rate of the G-mode phonons for both HOPG and SWNTs. The table presents the energies, $\hbar\omega_1$ and $\hbar\omega_2$, of the daughter phonons into which the G-mode phonons decay, and the decay rate, $\Gamma_0$, extrapolated to zero temperature.

| Material | $\hbar\omega_1$ [cm$^{-1}$] | $\hbar\omega_2$ [cm$^{-1}$] | $\Gamma_0$ [cm$^{-1}$] | $\Gamma_0$ [ps$^{-1}$] |
|---|---|---|---|---|
| HOPG | 535 ± 190 | 1050 ± 190 | 2.5 ± 0.3 | 0.47 ± 0.06 |
| SWNTs | 330 ± 41 | 1255 ± 40 | 4.2 ± 0.3 | 0.80 ± 0.05 |

# References

∗ Current address: Department of Physics and Astronomy, Iowa State University and Ames Laboratory-USDOE, Ames, IA 50011, USA; ioannis@iastate.edu